\documentstyle[psfig]{article}

\makeatletter

\def\enum{\ifnum \@enumdepth >3 \@toodeep\else
        \advance\@enumdepth \@ne 
        \edef\@enumctr{enum\romannumeral\the\@enumdepth}\list
        {\csname label\@enumctr\endcsname}
        {\setlength{\topsep}{1mm}
        \setlength{\parsep}{0mm}
        \setlength{\itemsep}{0mm}
        \setlength{\labelsep}{2mm}
        \settowidth{\leftmargin}{M.}
        \addtolength{\leftmargin}{\labelsep}
        \usecounter{\@enumctr}
        \def\makelabel##1{\hss\llap{##1}}}\fi}

\def\itemiz{\ifnum \@itemdepth >3 \@toodeep\else \advance\@itemdepth \@ne
        \edef\@itemitem{labelitem\romannumeral\the\@itemdepth}%
        \list{\csname\@itemitem\endcsname}{
        \setlength{\topsep}{1mm}
        \setlength{\itemsep}{1mm}
        \setlength{\labelsep}{2mm}
        \settowidth{\leftmargin}{M}
        \addtolength{\leftmargin}{\labelsep}
        \def\makelabel##1{\hss\llap{##1}}}\fi}

\def\thebibliography#1{\section*{References\@mkboth
 {REFERENCES}{REFERENCES}}\list
 {[\arabic{enumi}]}{\settowidth\labelwidth{[#1]}\leftmargin\labelwidth
 \setlength{\parsep}{0mm}
 \setlength{\itemsep}{0mm}
 \advance\leftmargin\labelsep
 \usecounter{enumi}}
 \def\newblock{\hskip .11em plus .33em minus .07em}
 \sloppy\clubpenalty4000\widowpenalty4000
 \sfcode`\.=1000\relax}

\def\section{\@startsection {section}{1}{\z@}{-3.5ex plus -1ex minus 
 -.2ex}{2.3ex plus .2ex}{\large\bf}}
 
\def\@opargbegintheorem#1#2#3{\it \trivlist
      \item[\hskip \labelsep{{\bf #1\ #2\ }{\rm #3}}]}

\makeatother

\def\bibtitle#1#2{#1}					
\def\bibref#1#2#3#4#5{#1 {\bf #2}:#3 (#5)}		
\def\bibarticle#1#2#3#4#5#6#7{\bibtitle{#1}{#2}, \bibref{#3}{#4}{#5}{#6}{#7}.}
\def\bibpreprint#1#2#3#4{#1, preprint {\tt #3} (#4).}
\def\bibbook#1#2#3#4{#1, {\em #2}, (#3, #4).}

\def\JPA{J. Phys. A}

\def\JSP{J. Stat. Phys.}

\def\PD{Physica D}
\def\PL{Phys. Letters}

\def\PRL{Phys. Rev. Letters}
\def\RMP{Rev. Mod. Phys.}

\def\dateen{\the\day\space\ifcase\month\or January \or February \or March \or
April \or May \or June \or July \or August \or September
\or October \or November \or December \fi\space\the\year}

\def\writefig#1 #2 #3 {\rlap{\kern #1 truecm
\raise #2 truecm \hbox{#3}}}
\def\figtext#1{\smash{\hbox{#1}}
\vspace{-5mm}}

\def\math#1{\ifmmode
\mathchoice{\mbox{$\displaystyle\rm#1$}}
{\mbox{$\textstyle\rm#1$}}
{\mbox{$\scriptstyle\rm#1$}}
{\mbox{$\scriptscriptstyle\rm#1$}}\else
{\mbox{$\rm#1$}}\fi}

\def\vec#1{\ifmmode
\mathchoice{\mbox{$\displaystyle\bf#1$}}
{\mbox{$\textstyle\bf#1$}}
{\mbox{$\scriptstyle\bf#1$}}
{\mbox{$\scriptscriptstyle\bf#1$}}\else
{\mbox{$\bf#1$}}\fi}

\newtheorem{theorem}{Theorem}
\newtheorem{proposition}{Proposition}
\newtheorem{lemma}{Lemma}
\newtheorem{definition}{Definition}
\newtheorem{corollary}{Corollary}

\def\bbbz{{\mathchoice {\hbox{$\sf\textstyle Z\kern-0.4em Z$}}
{\hbox{$\sf\textstyle Z\kern-0.4em Z$}}
{\hbox{$\sf\scriptstyle Z\kern-0.3em Z$}}
{\hbox{$\sf\scriptscriptstyle Z\kern-0.2em Z$}}}}

\def\bbbr{{\rm I\!R}} 

\def\bbbc{{\mathchoice {\setbox0=\hbox{$\displaystyle\rm C$}\hbox{\hbox
to0pt{\kern0.4\wd0\vrule height0.9\ht0\hss}\box0}}
{\setbox0=\hbox{$\textstyle\rm C$}\hbox{\hbox
to0pt{\kern0.4\wd0\vrule height0.9\ht0\hss}\box0}}
{\setbox0=\hbox{$\scriptstyle\rm C$}\hbox{\hbox
to0pt{\kern0.4\wd0\vrule height0.9\ht0\hss}\box0}}
{\setbox0=\hbox{$\scriptscriptstyle\rm C$}\hbox{\hbox
to0pt{\kern0.4\wd0\vrule height0.9\ht0\hss}\box0}}}}

\def\dis{\displaystyle}

\def\mysection#1{\vspace{2mm}\noindent{\bf #1}}

\def\C#1{{\cal C}^{#1}}
\def\pscal#1#2{#1\cdot #2}
\def\dpart#1#2{\frac{\partial #1}{\partial #2}}
\def\d#1{\math{d} #1}
\def\abs#1{\left|#1\right|}
\def\Order#1{{\cal O}(#1)}
\def\matrix22#1#2#3#4{\left(\begin{array}{ccc}
#1&\;&#2\\#3&\;&#4\end{array}\right)}
\def\dsmatrix22#1#2#3#4{\left(\begin{array}{cc}
\dis#1&\dis#2\\ & \\ \dis#3&\dis#4\end{array}\right)}

\def\brak#1{\left[#1\right]}

\def\eps{\varepsilon}
\def\e{\math{e}}
\def\d{\math{d}}

\def\dQ{\partial Q}
\def\half{\frac{1}{2}}
\def\Tr{\math{Tr}}
\def\rmin{\rho_{\math{min}}}
\def\rmax{\rho_{\math{max}}}
\def\th{\theta}
\def\E{\abs{{\cal E}}}
\def\Ev{{\cal E}}
\def\dpsi{\Delta\psi}
\def\Arccos{\math{Arccos}}

\begin{document}

\title{Classical Billiards in a Magnetic Field and a 
Potential\thanks{Proceedings of the $3^{\math{rd}}$ 
	International Summer School/Conference in Maribor, 
	Slovenia, 24 June -- 6 July, 1996.}}
\author{N. Berglund\\
Institut de Physique Th\'eorique\\
EPFL, PHB--Ecublens\\
CH-1015 Lausanne, Switzerland}
\date{9 December 1996}

\maketitle

\begin{abstract}
We study billiards in plane domains, with a perpendicular magnetic 
field and a potential. We give some results on periodic orbits, KAM tori 
and adiabatic invariants. We also prove the existence of bound states 
in a related scattering problem.
\end{abstract}

Classical billiards are popular models for various physical systems, 
in fields ranging from mechanics of systems with impacts and 
ergodic theory \cite{KT} to semiclassical methods in quantum chaos. In 
particular, billiards in a magnetic field appear to be relevant 
for the study of transport properties in mesoscopic systems, 
diamagnetism and the quantum Hall effect (see for instance 
\cite{Trugman}).

In this work, we consider the classical motion of a charged particle 
in a plane domain, with a perpendicular magnetic field of intensity 
$B$ and a potential $V(\vec{x})$. We first discuss a method for 
proving existence of periodic and quasiperiodic orbits. Then we give 
some results on KAM tori and adiabatic invariants for billiards in a 
magnetic field. Finally, we consider the scattering on a hard disc 
in crossed electromagnetic fields, where we prove the existence of 
bound states. Details on the present results can be found in 
\cite{BK,BHHP,B}.


\section{Billiards and Periodic Orbits}

\mysection{Bouncing Map:}
We consider the classical motion of a particle in a connected domain $Q$ 
(which may be unbounded or not simply connected). The boundary $\dQ$ is 
parametrized by its arclength, $\vec{x}(s)=(X(s),Y(s))$, with 
$X'(s)^2+Y'(s)^2=1$; the unit tangent vector and the curvature are 
given respectively by $\vec{t}(s)=(X'(s),Y'(s))$ 
and $\kappa(s)=X'(s)Y''(s)-X''(s)Y'(s)$.

Inside $Q$, the billiard flow is defined by the Lagrangian
\begin{equation}
{\cal L}(\vec{x},\dot{\vec{x}}) = \half m \dot{\vec{x}}^2 + 
q \pscal{\dot{\vec{x}}}{\vec{A}(\vec{x})} - V(\vec{x}),
\label{defL}
\end{equation}
where $\vec{A}(\vec{x})=\half B(-y,x)$ is the vector potential in 
symmetric gauge (we will adopt the sign convention $qB<0$).

The dynamics is defined in the following way. Assume that the billiard 
particle starts on the boundary at $\vec{x}(s_0)$, with a velocity 
$\dot{\vec{x}}_0$ making an angle $\theta_0$ with $\vec{t}(s_0)$. It then 
evolves in $Q$ according to the Lagrange equations. If the particle 
returns to the boundary, at a point $\vec{x}(s_1)$, with a velocity 
$\dot{\vec{x}}_1$ making an angle $-\theta_1$ with $\vec{t}(s_1)$, it is 
reflected elastically, meaning that it leaves the boundary again with 
an angle $\theta_1$ (Fig.\ 1). 
The $i$--th collision may thus be parametrized by the coordinates 
$(s_i,\theta_i)$, or, alternatively, by the {\em Birkhoff variables} 
$(s_i,u_i)\equiv z_i$, where  
$u_i=\dot{\vec{x}}_i\cdot\vec{t}(s_i)=\abs{\dot{\vec{x}}_i}\cos\theta_i$ 
denotes the tangent velocity.

\begin{figure}
 \centerline{\psfig{figure=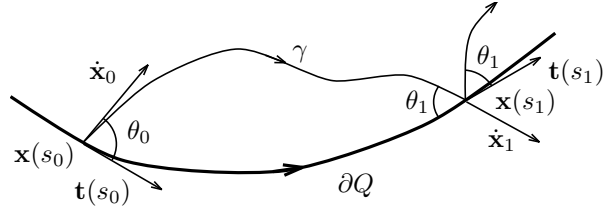,height=25mm,clip=t}}
 \figtext{
 	\writefig	8.1	0.4	$\partial Q$
 	\writefig	7.5	2.2	$\gamma$
 	\writefig	5.3	1.1	$\theta_0$
 	\writefig	9.0	1.5	$\theta_1$
 	\writefig	9.95	2.1	$\theta_1$
 	\writefig	3.8	0.8	$\vec{x}(s_0)$
 	\writefig	10.2	1.5	$\vec{x}(s_1)$
 	\writefig	4.6	0.3	$\vec{t}(s_0)$
 	\writefig	10.9	1.9	$\vec{t}(s_1)$
 	\writefig	4.8	1.9	$\dot{\vec{x}}_0$
 	\writefig	10.1	1.0	$\dot{\vec{x}}_1$
 }
 \vspace{3mm}
 \caption[]{The abscissa $s_0$ of the starting point and the angle 
 $\theta_0$ uniquely define the trajectory $\gamma$, and, if the 
 boundary $\partial Q$ is reached again, the coordinates 
 $(s_1,\theta_1)$ of the next collision.}
\end{figure}

As long as the particle returns to the boundary, 
we may thus describe the dynamics by the {\em bouncing map} 
$T: z_0\mapsto z_1$. 

\mysection{Generating function:}
A very useful tool to study the bouncing map is its {\em generating 
function} $G(s_0,s_1)$ \cite{Meiss}, which satisfies
\begin{equation}
\dpart{G}{s_0} = -u_0,\qquad \dpart{G}{s_1} = u_1.
\label{defG}
\end{equation}
It can be constructed in the following way: let $\gamma$ be a trajectory 
connecting the points $\vec{x}(s_0)$ and $\vec{x}(s_1)$, and 
$F(s_0,s_1)=\int_\gamma \pscal{\vec{p}}{\d\vec{x}}$ the action along 
$\gamma$, where $\vec{p} = \partial_{\dot{\vec{x}}}{\cal L} = 
m\dot{\vec{x}}+q\vec{A}(\vec{x})$ is the momentum. We know from analytical 
mechanics that for infinitesimal variations of the end points
$\d \vec{x}_i=\vec{t}(s_i)\d s_i$, the change of action is 
$\d F=-\pscal{\vec{p}_0}{\d \vec{x}_0} + 
\pscal{\vec{p}_1}{\d \vec{x}_1}$.
It is then easy to check that 
\begin{equation}
G(s_0,s_1) = \frac{1}{m}F(s_0,s_1) 
+ \frac{qB}{2m}\int_{s_0}^{s_1} Y(s)X'(s) - X(s)Y'(s)\,\d s
\label{G}
\end{equation}
satisfies the relations (\ref{defG}).

Here we have assumed that there is a unique trajectory connecting the 
two points on the boundary. In fact, there may be several or no solution 
of the Lagrange equations for given end points, so that the 
generating function may be multiply defined on some domain, and not exist 
on another one. This gives rise to new complications, but the main 
arguments presented in the following can be transposed to this more 
difficult situation.

\mysection{Periodic orbits:}
The relation (\ref{defG}) is useful to compute periodic orbits. If 
$s_0, s_1, \ldots s_{n-1}$ is a sequence of arclengths on the boundary, 
we define the {\em $n$--point generating function}
\begin{equation}
G^{(n)}(s_0, s_1, \ldots s_{n-1}) = G(s_0,s_1) + G(s_1,s_2) + 
\cdots + G(s_{n-1},s_0).
\label{def_Gn}
\end{equation}
The law of specular reflection implies that there is a periodic orbit 
of period $n$, 
hitting the boundary at $\vec{x}(s_0), \ldots, \vec{x}(s_{n-1})$, if and 
only if $\partial G^{(n)}/\partial s_i = 0$ 
for each $i$ (assuming $G^{(n)}$ is defined and sufficiently differentiable); 
in other words, periodic orbits correspond to stationary points of $G^{(n)}$.

The linear stability of the orbit is determined by noting that 
\begin{equation}
\begin{array}{c}
\d u_0 = - G_{20} \d s_0 - G_{11} \d s_1 \\
\d u_1 = \phantom{-} G_{11} \d s_0 + G_{02} \d s_1 
\end{array}
\qquad
G_{jk} = \dpart{^{j+k}G}{s_0^j\,\partial s_1^k}(s_0, s_1),
\label{lin1}
\end{equation}
implies $\d z_1=T'(s_0,s_1) \d z_0$, where
\begin{equation}
T'(s_0, s_1) = -\frac{1}{G_{11}}\matrix22{G_{20}}{1}
{G_{20}G_{02} - G_{11}^2}{G_{02}}
\label{lin2}
\end{equation}
is the Jacobian matrix of the bouncing map.
Since $T'$ has unit determinant, $T$ is area--preserving.
For $n$ iterations, $\d z_n=S_n\d z_0$, where 
$S_n(s_0, \ldots, s_{n-1}) = T'(s_0, s_{n-1}) T'(s_{n-1}, s_{n-2}) 
\cdots T'(s_1, s_0)$. The eigenvalues of $S_n$ are 
$\lambda_\pm = t\pm\sqrt{t^2-1}$, where $t=\half\Tr S_n$. 
Hence, the orbit is hyperbolic if $\abs{t}>1$, elliptic if $\abs{t}<1$ 
(under certain conditions, $t$ may be related to second derivatives 
of $G^{(n)}$ \cite{MM}).

The center manifold theorem implies that hyperbolic orbits are unstable, 
even when nonlinear terms are taken into account. Elliptic orbits are 
generically nonlinearly stable (in the sense of Lyapunov), as a consequence 
of the KAM theorem. A standard result is

\begin{theorem}
\label{th1}
Let $T$ be measure--preserving and $\C{5}$ in a 
neighborhood of the periodic orbit. Assume that the eigenvalues 
are such that $(\lambda_\pm)^3\neq 1$ and $(\lambda_\pm)^4\neq 1$. 
There exists $C$, depending only on the second and third derivatives 
of $T$ along the orbit (and which can be explicitly computed), such that 
if the non--degeneracy condition $C\neq 0$ is satisfied, 
any point of the orbit has a neighborhood which is invariant under the 
map $T^n$.
\end{theorem}

This result is proved by applying Moser's theorem \cite{Mo} to the Birkhoff 
normal form of $T^n$.


\begin{figure}[b]
 \centerline{\psfig{figure=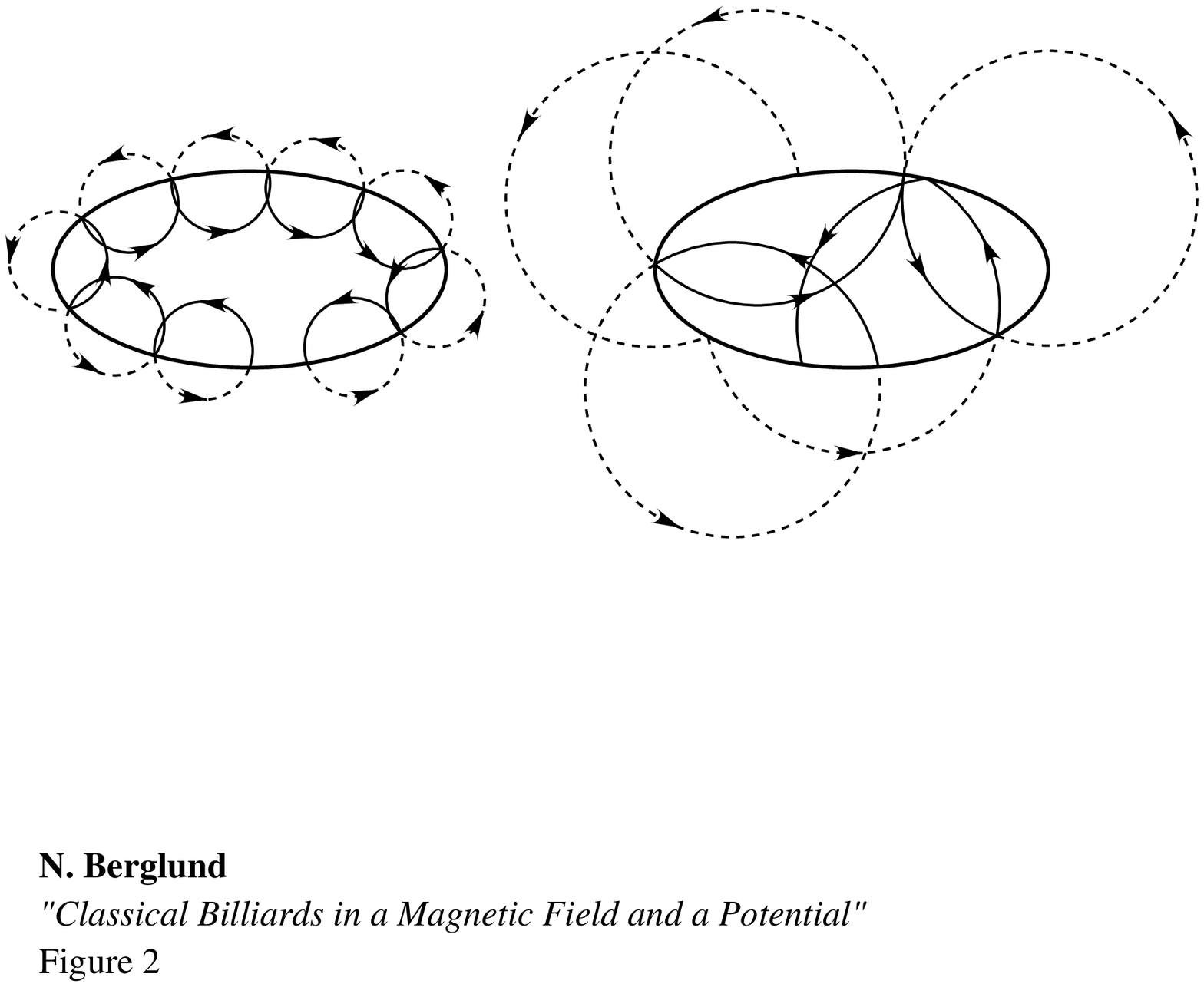,height=42mm,clip=t}}
 \caption[]{Inside--outside duality. In (a), the $\mu$--intersection 
 property is satisfied, there is a one--to--one correspondence between 
 inside and outside trajectories. If this property is not satisfied, 
 the duality may be destroyed (b).}
 \figtext{
 	\writefig	2.5	6.0	a
 	\writefig	7.0	6.0	b
 }
\end{figure}

\section{Billiards in a Magnetic Field}

The particular case when there is only the magnetic field 
($V(\vec{x})=0$) was first considered by Robnik and Berry \cite{RB}. 
For a given energy $E$, the trajectories are arcs of Larmor 
radius $\mu=\sqrt{2mE}/\abs{qB}$. The generating function and the Jacobian 
matrix can be explicitly expressed in terms of geometric properties of the 
boundary \cite{BK}.

We first consider the billiard in a convex domain, i.e., we assume that the 
radius of curvature $\rho(s) = 1/\kappa(s)$ is smooth and bounded by 
positive constants $\rmin$ and $\rmax$. 
An important class of orbits are the ``whispering gallery modes'': 
they correspond physically to quasiperiodic trajectories skipping along the 
boundary; in phase space, these solutions live on invariant curves which are 
close to $\th = 0, \pi$. In the zero field case, existence of such orbits was 
proved by Lazutkin \cite{L}. 
As remarked in \cite{RB}, when a magnetic field is added, the 
dynamics near the boundary depend on the relative value of $\mu$, 
$\rmin$ and $\rmax$. This is confirmed by the following result:

\begin{theorem}[\cite{BK}]
\label{th2}
Assume the boundary is $\C{5}$ and satisfies 
$0<\rmin\leq\rho(s)\leq\rmax\leq\infty$. There exists a Cantor set 
of invariant curves with positive measure in the three following 
situations:
\begin{enum}
\item	For $0<\mu<\infty$, near $\theta=\pi$. It corresponds to 
	backward skipping trajectories, which are curved towards 
	the boundary.
	
\item	For $\mu>\rmax$, near $\theta=0$. 
	It corresponds to forward skipping trajectories which are 
	curved away from the boundary.
	
\item	For $\mu<\rmin$, near $\theta=0$. It corresponds to 
	backward skipping trajectories starting with a forward glancing 
	velocity, and performing almost complete Larmor circles.
\end{enum}
\end{theorem}

The proof relies on a perturbative expression of the bouncing map for 
small $\sin\theta$, which can by analyzed by Moser's theorem \cite{Mo}. The 
difference between the regimes $\theta\sim 0$ and $\theta\sim \pi$ is due 
to the symmetry breaking effect of the magnetic field. 
When $\rmin<\mu<\rmax$, invariant curves near $\theta=0$ are absent due to 
discontinuities by tangency. 
Note that in contrast with a theorem by Mather \cite{Ma}, in a magnetic 
field invariant curves still exist when the curvature is allowed to vanish. 
In this respect, the magnetic field has a stabilyzing effect.

This result can be extended to more general billiard domains. Consider 
for instance the billiard {\em outside} a given convex curve. One can 
try to construct a trajectory of the outside billiard by completing 
every arc of an inside trajectory to a full circle (Fig.\ 2). 
There is a one--to--one correspondence between inside and outside 
orbits if the following property is satisfied:

\begin{definition}
A closed plane $\C{2}$ curve is said to have the {\em $\mu$--intersection 
property} for some $\mu>0$ if any circle of radius $\mu$ intersects it 
at most twice.
\end{definition}

\begin{lemma}[\cite{BK}]
A closed plane convex curve with extremal radii of curvature $\rmin$ and 
$\rmax$ satisfies the $\mu$--intersection property if $\mu<\rmin$ or 
$\mu>\rmax$.
\end{lemma}

We conlude that the inside and outside billiards are equivalent 
in low or high magnetic field. In fact, it is possible to show that the 
inside--outside duality remains true for backward skipping trajectories, 
even for intermediate magnetic field. Thus, Theorem \ref{th2} is valid 
for outside as well as for inside billiards, implying that this large 
class of billiards is not ergodic.

More generally, one can consider domains which are not convex, but whose 
boundary has a bounded curvature: $\abs{\kappa(s)}\leq1/\rmin$. It is 
possible to show that point 1.\ of Theorem \ref{th2} remains true if 
$\mu<\rmin$, which is once again a manifestation of the stabilyzing 
effect of the magnetic field. 
In this case, it is of particular interest to consider the strong magnetic 
field limit.

\begin{proposition}[\cite{BK}]
Assume the boundary is $\C{k}$, $k\geq 3$, and has a bounded curvature. 
For sufficiently small $\mu$, the bouncing map is $\C{k-1}$ and takes 
the form
\begin{eqnarray}
s_1 &=& s_0 - 2\mu\sin\th_0 + \mu^2\sin\th_0 \, a(s_0,\th_0,\mu)
\pmod{\abs{\dQ}},
\nonumber \\
\th_1 &=& \th_0 + \mu^2 \sin^2\th_0 \, b(s_0,\th_0,\mu).
\label{Tmu}
\end{eqnarray}
The functions $a\in \C{k-2}$ and $b\in \C{k-3}$ are uniformly bounded for 
$s\in\bbbr,0\leq\th\leq\pi$, $\abs{\dQ}$-periodic in $s$, and admit
expansions in $\mu$ which can be explicitly computed.
\end{proposition}

The bouncing map (\ref{Tmu}) has the structure of a perturbed integrable 
map, where the factors $\sin\th_0$ ensure that the boundaries $\th=0,\pi$ 
are fixed. One can again study invariant curves, of the 
form $I(s,\th) = \mbox{const}$, using 
Moser's theorem, although the analysis is complicated by the fact that 
the frequency $\Omega(\th) = - 2\mu\sin\th$ is multiplied by the small 
parameter $\mu$ and is not monotonic.

An alternative approach to KAM theory is to construct {\em adiabatic 
invariants} $J(s,\th)$, such that $J(s_1,\th_1)-J(s_0,\th_0)$ 
is as small as possible.

\begin{theorem}[\cite{BK}]
\label{th3}
If the boundary $\dQ$ is $\C{k}$, $k\geq 3$, and has bounded curvature, 
there exists a function $J(s,\th)$ such that 
$J(s_1,\th_1)=J(s_0,\th_0)+\Order{\mu^{k+1}}$.
If $\dQ$ is analytic, there exists a function $J(s,\th)$ such that 
$J(s_1,\th_1)=J(s_0,\th_0)+\Order{\e^{-1/C\abs{\mu}}}$.
\end{theorem}

In fact, Theorem \ref{th3} is true for a large class of maps, including 
(\ref{Tmu}). In the case of a billiard, 
\begin{equation}
J(s,\th;\mu) = \th + 
\mu\sin\th\brak{\frac{1}{3}\kappa(s) + \frac{2}{9}\mu\cos\th\kappa(s)^2 
+ \Order{\mu^2}}.
\end{equation}
If the boundary is analytic, Theorem \ref{th3} implies that for {\em any} 
initial condition $(s_0,\th_0)$, $(s_n,\th_n)$ remains at a distance of 
order $\mu$ from the level curve $J(s,\th)=J(s_0,\th_0)$, during a time 
of order $\e^B$.


\section{A Scattering System}

We now consider the case where the potential is given by a uniform 
electric field, $V(\vec{x}) = -q\pscal{\Ev}{\vec{x}}$, $\Ev=(0,\E)$.
One can introduce dimensionless variables such that the Lagrangian 
becomes
\begin{equation}
{\cal L} = \half \dot{\vec{x}}^2 + \half (y\dot{x} - x\dot{y}) - \eps y,
\label{scat1}
\end{equation}
where $\eps=\abs{\Ev m/qB^2}$ measures the strength of the electric field.

The trajectories are cycloids of the form
\begin{equation}
x(\psi) = a + \eps\psi + \rho\cos(\psi-\bar{\psi}), \qquad
y(\psi) = b + \rho\sin(\psi-\bar{\psi}), 
\label{scat2}
\end{equation}
where $\rho$ reduces to the Larmor radius when $\eps=0$. 
To account for the conservation of energy $E=\half(\eps^2+2\eps b+\rho^2)$, 
we introduce a second parameter $\mu=\sqrt{2E-\eps^2}$, so that 
$\rho = \sqrt{\mu^2-2\eps b}$. The range of the tangent velocity is
$\abs{u}\leq\mu(1+\Order{\eps})$.

\begin{figure}
 \centerline{\psfig{figure=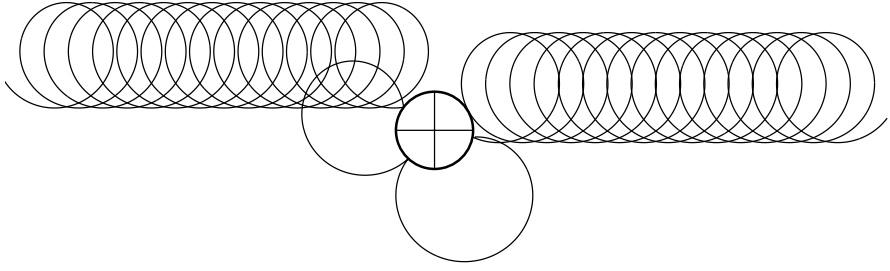,height=35mm,clip=t}}
 \caption[]{A trajectory scattered off the hard disc. Trajectories 
 coming in from infinity leave the scatterer again with probability 
 one. However, some orbits may form ``bound states'' which 
 are indefinitely bouncing on the scatterer.}
\end{figure}

In \cite{BHHP}, the following problem was studied: consider the billiard 
outside a circular scatterer, parametrized by $\vec{x}(s)=(\cos s,\sin s)$.
Are there trajectories which remain ``trapped'' in the vicinity of the 
scatterer, in spite of the drift due to the electric field? In fact, one 
can easily show that a particle drifting in from infinity will leave the 
scatterer again with probability one (Fig.\ 3). However, it is possible that 
some trajectories bounce on the scatterer indefinitely in the past and in the 
future, forming a classical ``bound state''. To show this, we first need 
to construct the generating function. Given two end points $\vec{x}(s_0)$ 
and $\vec{x}(s_1)$, we have to determine the equation of the corresponding 
cycloid (\ref{scat2}), and use the general formula (\ref{G}). 
The perturbative result is:

\begin{proposition}[\cite{B}]
Let $s_\pm=(s_1\pm s_0)/2$.
There are positive constants $c_1$, $c_2$ and $\eps_0$, such that 
for $c_1\eps < s_- < \pi-c_1\eps$, $\mu > 1 + c_2\eps$ and 
$0 \leq \eps < \eps_0$, 
the generating function of the bouncing map is unique, an analytic function 
of $s_\pm$, $\mu$ and $\eps$, and admits the expansion
\begin{eqnarray}
G(s_-, s_+) &=& s_- + \half \mu^2 \dpsi - (C+R)S - 
\eps\left[2S+(C+R)\dpsi\right] \sin s_+  \label{genscat} \\
& & +  \eps^2\left[\left(\frac{(C+R)^2S}{\mu^2R} + \frac{C+R}{R}\dpsi 
+ \frac{\mu^2}{4RS}\dpsi^2\right) \sin^2 s_+ - \frac{R}{4S}\dpsi^2 
\right] + \Order{\eps^3},\nonumber
\end{eqnarray}
where $C$, $S$, $R$, $\dpsi$ denote functions of $s_-$ alone:
\begin{equation}
C = \cos s_-,\quad S = \sin s_-, \quad R = \sqrt{\mu^2 - S^2}, \quad
\dpsi = 2\pi - \Arccos [1-2 S^2/\mu^2].
\label{gen2}
\end{equation}
\end{proposition}

The point is that the generating function has the form 
$G(s_0,s_1,\eps) = G_0(s_1-s_0) + \eps G_1(s_0,s_1,\eps)$. 
If we substitute this expression in (\ref{defG}), we obtain that 
the bouncing map has the structure of a perturbed integrable map:

\begin{corollary}
There is a positive $c_3$ such that for $\abs{u}<\mu(1-c_3\eps)$, 
$\mu > 1 + c_2\eps$ and $0 \leq \eps < \eps_0$, the trajectory 
returns to the boundary, the bouncing 
map is analytic and of the form
\begin{eqnarray}
s_1 &=& s_0 + \Omega(u_0) + \eps f(s_0,u_0,\eps)
\pmod{2\pi},
\nonumber \\
u_1 &=& u_0 + \eps g(s_0,u_0,\eps).
\label{Tscat}
\end{eqnarray}
\end{corollary}

This result asserts that the particle will return to the scatterer if 
it starts with a sufficiently large normal velocity. The problem is 
now to show that some of these trajectories indefinitely return to the 
scatterer. This can be achieved once again by using Moser's theorem: 
indeed the existence of two distinct invariant curves in phase space 
implies the region between them to be invariant under the map.

\begin{theorem}[\cite{B}]
There is a positive $\eps_1$ such that for $\eps<\eps_1$ and 
$\mu>1+c_2\eps$, the scattering system has a set of bound states with 
positive measure.
\end{theorem}

\begin{figure}
 \centerline{\psfig{figure=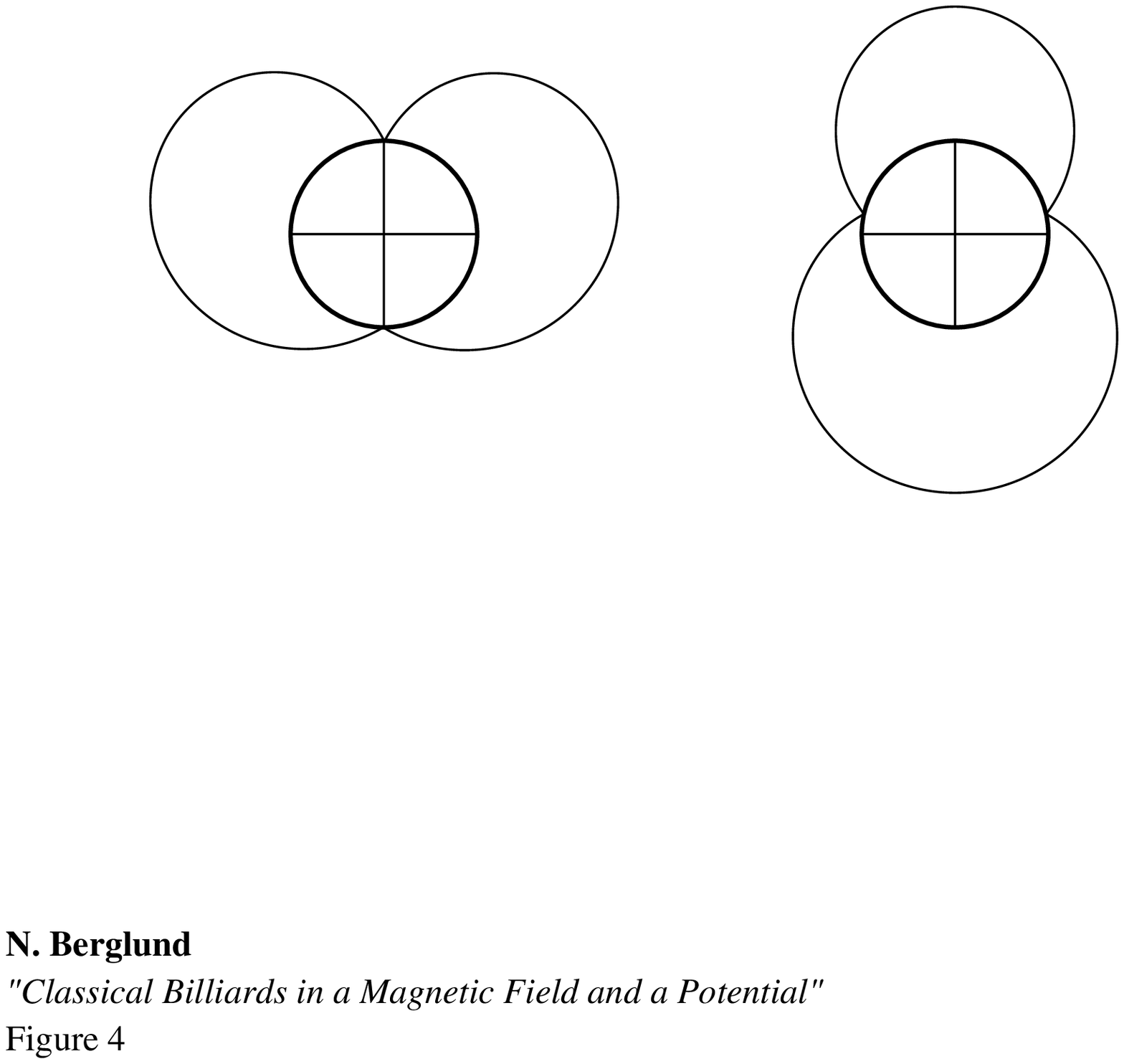,height=35mm,clip=t}}
 \caption[]{Orbits of period 2, for $\eps=0.1$ and $\mu=1.5$: 
 (a) hyperbolic orbit, (b) elliptic orbit. The elliptic orbit 
 is generically surrounded by a set of bound states.}
 \figtext{
 	\writefig	3.5	4.5	a
 	\writefig	8.0	4.5	b
 }
\end{figure}

The problem with KAM theory is that one has in general very bad 
estimates on the bound $\eps_1$. One can improve them by studying 
periodic orbits, which can be surrounded by islands 
of stability for much higher values of the electric field. As described 
in Section 1, we can compute the two--point function $G^{(2)}$ and 
analyze its stationnary points (for this purpose, we needed to 
know (\ref{genscat}) at second order in $\eps$). 
In this way we can prove the existence of two orbits of period 2:
\begin{enum}
\item 	An orbit hitting the scatterer at $s=\pi/2, 3\pi/2$, which is 
	hyperbolic for small $\eps$ (Fig.\ 4a).
	
\item	An orbit hitting the scatterer at $s=\delta, \pi-\delta$, 
	where
	$
	\delta = \eps\dpsi(\pi/2)/R(\pi/2) + \Order{\eps^3},
	$
	which is elliptic for $0<\eps<\eps_2$ (Fig.\ 4b). 
	Using Theorem \ref{th1}, 
	we can show that if the orbit is elliptic, then it is stable 
	with probability 1 with respect to $\d\mu\d\eps$. 
	We have the estimation
	\begin{equation}
	\eps_2\simeq\brak{\frac{4}{\mu^2}+
	\frac{4\dpsi(\pi/2)}{R(\pi/2)}+
	\dpsi(\pi/2)^2}^{-1/2},
	\end{equation}
	which is in good agreement with numerics (roughly, bound states 
	exist as long as the drift per cycle is smaller than the radius 
	of the scatterer).
\end{enum}

This method of searching elliptic orbits can be used to study existence 
of bound states for more general scatterers. In fact, if the billiard 
posseses elliptic orbits in zero electric field, they will generically 
survive small perturbations. 
Interesting open problems include (1) the existence of a critical 
electric field beyond which there are {\em no} bound states, and 
(2) an understanding of transport in phase space, and its influence 
on the transit time of a particle drifting in from infinity.


\section*{Acknowledgements}

Much of the work presented here is the result of fruitful collaborations 
with H.\ Kunz and with A.\ Hansen, E.\ H.\ Hauge, J.\ Piasecki. I would 
like to thank the organizers and participants of the $3^{\math{rd}}$ 
International Summer School/Conference in Maribor, Slovenia, for their 
interest. This work is supported by the Fonds National Suisse de la 
Recherche Scientifique.


\end{document}